\documentclass{ws-procs975x65}

\begin{document}

\title{MANY-BODY METHODS FOR NUCLEAR SYSTEMS AT SUBNUCLEAR DENSITIES}

\author{Armen~Sedrakian}
\address{Institute for Theoretical Physics, J. W. Goethe-Universit\"at,\\
D-60054 Frankfurt am Main, Germany\\
$^*$E-mail: sedrakian@th.physik.uni-frankfurt.de\\
}

\author{John W.~Clark}
\address{Department of Physics, Washington University,\\ 
St. Louis, Missouri 63130, USA\\
$^*$E-mail: jwc@wuphys.wustl.edu}

\begin{abstract}
%
This article provides a concise review of selected topics in
the many-body physics of low density nuclear systems. The discussion 
includes the condensation of alpha particles in supernova envelopes, 
formation of three-body bound states and the BEC-BCS crossover in dilute 
nuclear matter, and neutrino production in $S$-wave paired superfluid
neutron matter.
\end{abstract}


\bodymatter

\section{Introduction}\label{sec:Intro}

The physics of matter at subnuclear densities $\rho \in [10^{11} - 10^{14}]$
g cm$^{-3}$ is of great interest for the astrophysics of compact objects. The 
``hot'' stage of evolution of matter, in which temperatures are in the 
range of tens of MeV, is associated with the dynamics of supernova explosions. 
Knowledge of the equation of state, composition, and weak-interaction 
processes are of prime importance for an understanding the mechanism 
of explosion, the formation of neutrino spectra at the neutrinosphere, 
and the elemental abundances of the low-density matter in the supernova 
winds that are prerequisite for the onset of $r$-process nucleosynthesis.  
Days to weeks after the supernova explosion subnuclear matter has 
become  ``cold'', with temperatures $T  <0.1$ MeV. 
Moreover,
the properties of the subnuclear matter forming the crust of a neutron star 
are of fundamental importance for the entire spectrum of observable
manifestations of pulsars, ranging for example from superfluid rotation
dynamics to magnetic field evolution to neutrino cooling.
 
Nuclear matter at subnuclear densities is a strongly correlated 
system in which the relevant degrees of freedom are well established and the 
interactions are constrained by experiment. The challenge lies in the 
many-body treatment of this system where macroscopic quantum phenomena such
as Bose-Einstein condensation of deuterons and alpha particles exist 
as well as the BCS pairing in neutron matter. Our aim here is to 
describe some of the many-body methods for dealing with such correlated 
states of matter.  We will pay less attention to the physical setting 
and implications of the results; the reader concerned with these 
issues is referred to the original literature cited among the
references.

\section{Bose-Einstein condensation: a lattice Monte-Carlo perspective}

In this section we describe an approach to interacting Bose systems 
which is valid in the vicinity of the critical temperature of 
Bose-Einstein condensation (BEC). The method was put forward 
in the context of dilute gases interacting via {\it repulsive} 
two-body forces~\cite{Baym:1999ws,ZinnJustin:2000dr} and has 
since been reformulated for a strongly correlated system interacting 
with {\it attractive} two-body and repulsive three-body 
forces~\cite{Sedrakian:2004fh}. The method has been applied 
to Bose condensation of alpha particles in infinite matter. 
(Alternative studies are based on hypernetted-chain 
summations~\cite{ALPHA}.) Consider a  uniform, non-relativistic 
system of identical bosons described by the Hamiltonian
\begin{equation}\label{eq:1}
H = \int\!d^3x \Biggl[ \frac{\hbar^2}{2m}
     \mathbf\nabla\psi^{\dagger}({\mathbf x})\mathbf\nabla\psi({\mathbf x})
  -  \mu\vert\psi({\mathbf x})\vert^2
  +g_2\vert\psi({\mathbf x})\vert^4
  +  g_3\vert\psi({\mathbf x})\vert^6
\Biggr],\nonumber\\
\end{equation}
where $m$ is the alpha-particle mass, $\mu$ is the chemical potential, and
$\psi$ is the boson field.  Below, we shall implement lattice regularization.
The theory defined by Eq.~(\ref{eq:1}) can be mapped onto an effective 
scalar field theory within the finite-temperature Matsubara formalism.
Consider the fields $\psi$ and $\psi^{\dagger}$ as periodic 
functions of the imaginary time $\tau\in [-\beta, \beta]$, where
$\beta = 1/T$ is the inverse temperature. Next, decompose the fields into 
discrete Fourier series
\begin{equation}\label{FOURIER}
\psi({\mathbf x},\omega_{\nu}) =
\sum_{\nu=-\infty}^{\infty}e^{i\omega_{\nu}\tau}\psi({\mathbf x},\tau)
 = \psi_0({\mathbf x})
 +\sum_{\nu=-\infty,~\nu\neq 0}^{\infty}e^{i\omega_{\nu}\tau}\psi({\mathbf x},\tau),
\end{equation}
where the Fourier frequencies $\omega_{\nu}$ are the bosonic Matsubara
modes $\omega_{\nu} = 2\pi i \nu T$ (with $\nu$ taking integer values).
The Matsubara Green's function is given by 
$
{G}^M(\omega_{\nu},{\mathbf x}) = [i\omega_{\nu} -(2m)^{-1}
\mathbf\nabla^2
+\mu]^{-1}.
$
Here the chemical potential may include any contribution from 
the momentum- and energy-independent part of the self-energy; we also assume
that any momentum and energy dependent parts are absorbed in the mass and 
the wave-function renormalizations, respectively. 
Since $\mu\to 0$ near $T_c$, 
the characteristic scales of spatial variations of the Green's function 
with non-zero Matsubara frequencies are $l = (2m\omega_{\nu})^{-1/2}$, 
which are of the order of the thermal wave-length $\lambda = 
(2\pi/mT)^{1/2}$. The contribution of the non-zero Matsubara modes to 
the sum in Eq.~(\ref{FOURIER}) will be neglected since we are interested 
in scales $L\gg l$, which are characterized only by the zero-frequency 
modes. In terms of new real scalar fields $\phi_1$ and $\phi_2$ 
defined via the relations $\psi_0 = \eta(\phi_1+i\phi_2)$ and 
$\psi^{\dagger}_0 = \eta(\phi_1-i\phi_2)$, where 
$\eta = \sqrt{m/\hbar^2\beta}$, the continuum action of the 
theory is given by
\begin{equation} \label{eq:6}
{S}\left(\phi\right)
 = \int d^3 x   \Biggl\{
\frac{1}{2}\sum_{\nu}\left[\partial_{\nu}\phi({\mathbf x})\right]^2
+\frac{r}{2}\phi({\mathbf x})^2
-\frac{u}{4!}\left[\phi({\mathbf x})2\right]^2
+\frac{w}{6!}\left[\phi({\mathbf x})2\right]^3
\Biggr\},
\end{equation}
where
$\phi^2 = \phi_1^2+\phi_2^2$, $r = -2 \beta\mu\eta^2$,
$u = 4! \beta g_2 \eta^4$ and $w = 6!\beta g_3 \eta^6$.
The action (\ref{eq:6}) describes a classical
$O(2)$ symmetric scalar $\phi^6$ field theory in 
three spatial dimensions (3D).
The positive sextic interaction guarantees that the
energy is bound from below, which would not otherwise be the case
because of the negative sign of the quartic term describing
the attractive two-body interactions.
The characteristic length scale of the theory is set
by the parameter $u$, which has the dimension of inverse length;
the dimensionless parameter of the lattice theory
is $ua_L$, where $a_L$ is the lattice spacing. The thermodynamic 
functions of the model are obtained from the partition function
\begin{equation} \label{eq:5}
{Z} = \int [d\phi({\mathbf x})] {\rm exp}
\left[-{S}\left(\phi\right)\right].
\end{equation}
For example, the expectation value of the particle number density
is given by
$n_{\alpha} = \langle \psi^*\psi\rangle = (\beta V)^{-1}
\partial {\rm ln}{Z} /\partial \mu$, where $V$ is the volume. The continuum
theory is now discretized on a lattice by replacing the integrations 
over spatial coordinates by a summation over lattice sites.
The discretized version of
the continuum action (\ref{eq:6}) is
\begin{equation} \label{eq:7}
{S}_L\left(\phi\right) = \sum_i \Biggl\{-2\kappa \sum_{\nu}
\phi_L({\mathbf x})\phi_L({\mathbf x}+a\hat\nu)\phi_L({\mathbf x})^2 
+ \lambda\left[1+\phi_L({\mathbf x})^2\right]^2
 \lambda+ \zeta\left[\phi_L({\mathbf x})^2\right]^3\Biggr\},\nonumber
\end{equation}
the $\nu$ summation being carried out over unit vectors in three 
spatial directions
(nearest neighbor summation).  The hopping parameter $\kappa $ 
and the two- and three-body coupling constants $\lambda$ and 
$\zeta$ are related to the parameters of the continuum action
through $a_L^2r = (1-2\lambda)/\kappa-6$,
$\lambda = a_L\kappa^2 u/6$, and $\zeta = w\kappa^3/90$.
The lattice and continuum fields are related by $\phi_L({\bf x})
= (2\kappa/a_L)^{1/2}\phi({\bf x})$.
The components of the spatial vector $x_{\nu}$ are discretized at 
integral multiples of
the lattice spacing $a_L$: $x_{\mu} = 0, a_L, \dots (L_{\nu}-1)a_L$.
For a simple cubic lattice in
3D
with periodic boundary
conditions imposed on the field variable, one has $\phi_L(x+a_LL)
= \phi_L(x)$ (N.B.\ for a box of length $L$
there are $L^3$ (real) variables within the volume $(La_L)^3$).

In Ref.~\citen{Sedrakian:2004fh},
the field configurations on the lattice were evolved using a
combination of the heatbath and local Metropolis algorithms,
executing $10^5-10^6$ equilibration sweeps 
for lattices sizes from $8^3$ to $64^3$.
Once the field values on the lattice were determined, these were
transformed into their counterparts in the continuum theory
to obtain the statistical average value $\langle H\rangle$ 
of the Hamiltonian, i.e., the grand canonical (thermodynamical)
potential $\Omega$ as a function of density. 
The critical temperature $T_c$ for Bose-Einstein condensation
can be obtained from the simulations~\cite{Sedrakian:2004fh}. 
In practice one
obtains the density $n(T)$, rather than $T(n)$ directly,
at constant fugacity $z\to 1$. Working at small fugacity
${\rm  log}~z= -0.1$, the density of the system at constant
small chemical potential can be computed and 
the associated temperature is then identified with
$T_c$. 

\section{From pair condensation to three-body bound states}

The few-body bound states are of interest in the ``hot'' stage 
of compact stars to the extent that they can provide a very efficient 
source of opacity for neutrinos propagating through matter. This
situation can be compared to that of radiative transfer in ordinary 
stars, where the photoabsorption on the weakly-bound negative ion of 
hydrogen (H$^-$) largely determines the opacity. If the system is 
isospin-symmetric, the pairing occurs in the $^3S_1-^3D_1$ 
partial-wave channel. There is a two-body bound state in this 
channel in free space -- the deuteron; hence the BCS to BEC
crossover arises in this new context~\cite{NSR,Alm:1991ne,Lombardo:2001ek}. 
Following up on the conjecture of Nozi\`eres and Schmitt-Rink~\cite{NSR}, 
one may attempt to describe this crossover within mean-field BCS theory. 
The central numerical problem then reduces to solution of the gap 
equation
\begin{equation}\label{GAP2}
\Delta_{l}(p) = -\int\frac{dp'p'^2}{(2\pi)^2}\sum_{l'=0,2}V_{ll'}^{3SD1}(p,p')
 \frac{\Delta_{l'}(p')}{\sqrt{E(p')^2+D(p')^2 }}
 \left[1 - 2f(E(p')) \right],
\end{equation}
where $D^2(k) \equiv (3/8\pi)[\Delta_0^2(k)+\Delta_2^2(k)]$
is the angle-averaged neutron-proton gap function,
$V^{3SD1}(p,p')$ is the interaction in the $^3S_1-^3D_1$ channel,
$E(p)$ is the quasiparticle spectrum, and $f$ is the Fermi distribution 
function.  The chemical potential is determined self-consistently from the
gap equation (\ref{GAP2}) and the expression for the density.

We now outline an algorithm for numerical solution of
the gap equation, which can be applied 
to arbitrary potentials that are attractive at
large separations.\cite{Sedrakian:2005db}
The method is effective in dealing with the hard
core (short-range repulsion) in nuclear potentials and could
be useful for other systems featuring short-range repulsive interactions.
The starting point is the gap equation with an ultraviolet momentum 
cutoff $\Lambda \ll \Lambda_P$, where $\Lambda_P$ is
of the order of the natural (soft) cutoff of the potential.
Successive iterations, which generate approximant $\Delta_i$ to
the gap function from approximant $\Delta^{(i-1)}$ ($i=1,2, \ldots$), 
are determined from
\begin{equation} \label{GAP3}
\Delta^{(i)} (p,\Lambda) = \int^{\Lambda}\frac{dp'p'^2}{(2\pi)3} 
V^{3SD1}(p,p')\frac{\Delta^{(i-1)}(p',\Lambda)}
{\sqrt{E(p')^2+D^{(i-1)}(p',\Lambda)^2}}\left[1 - 2f(E(p)) \right].
\end{equation}
The process is initialized by first solving Eq.~(\ref{GAP2}) for
$D(p_F)$, where $p_F$ is the Fermi momentum, assuming the gap function
to be a constant.  The initial approximant for the momentum-dependent gap 
function is then taken as $\Delta^{(i=0)} (p) = V(p_F,p)D(p_F)$. 
Two iteration loops are implemented at given chemical potential.
An internal loop operates at fixed 
$\Lambda$ and solves the gap equation (\ref{GAP3}) iteratively
for $i= 1,2,\dots$.  An external loop increments the cutoff 
$\Lambda$ until  $d\Delta(p,\Lambda)/d\Lambda \approx 0$.  The finite 
range of the potential guarantees that the external loop 
converges once the entire momentum range spanned by the potential 
is covered. Thus, choosing the starting $\Lambda$ small enough,
we execute the internal loop by inserting $\Delta^{(i-1)} (p)$ 
in the right-hand side of Eq.~(\ref{GAP3}) to obtain a new 
$\Delta_i (p)$ on the left-hand side, which in turn is re-inserted 
in the right-hand side.  This procedure 
converges rapidly to a momentum-dependent solution for the gap 
equation for $\Delta(p,\Lambda_j)\theta(\Lambda_j-p)$, where 
$\theta$ is the step function and the integer $j$ counts
the iterations in the external loop.

For the next iteration, the cutoff is incremented to $\Lambda_j =
\Lambda_{j-1}+\delta\Lambda$, where $\delta\Lambda\ll \Lambda_j$,
and the internal loop is iterated until convergence is reached.
The two-loop procedure is continued until $\Lambda_j > \Lambda_P$, 
after which the iteration is stopped, a final result for $\Delta(p)$ 
independent of the cutoff having been achieved.  Once this process
is complete, the chemical potential must be updated via the equation 
for the density.  Accordingly, a third loop of iterations 
seeks convergence between the the output gap function and the 
chemical potential, such that the starting density is reproduced. 

The BCS-BEC crossover~\cite{NSR} 
in nuclear systems has been studied as a function 
of density and asymmetry in the population of isospin states 
(proton-neutron asymmetry).  
A number of interesting features are 
revealed\cite{Lombardo:2001ek,Sedrakian:2006xm}. 
In the extreme low-density limit, the chemical potential changes 
its sign and tends to $-1.1$ MeV, half the deuteron binding 
energy. Thus, twice the chemical potential plays the role of 
the eigenvalue in the Schr\"odinger equation for two-body bound 
states in this, the BEC limit.  The pair function is very
broad in this low-density regime, indicating that the deuterons 
are well localized in space; conversely it is peaked in the 
BCS limit at high density where the Cooper pairs are correlated 
over large distances. 
In the case of asymmetric systems, the density distribution of 
the minority particles (protons) has a zero-occupation (blocking) 
region that is localized around their Fermi-surface.  Upon
crossover to the BEC side, the blocking region becomes wider 
and moves toward lower momenta.  Eventually there is 
a topological change in the Fermi surface: the states are 
occupied starting at some finite momentum and are empty below 
that point.  The Nozi\`eres--Schmitt-Rink conjecture~\cite{NSR} of
a smooth crossover from the BCS to the BEC limit does not 
hold in general for asymmetric systems; instead, phases 
with broken space symmetries intervene within a certain range of
population asymmetries (see Ref.~\citen{Sedrakian:2006xm} 
and references therein).

The three-body bound states can be computed from the three-body scattering 
matrix, which is written as ${\cal T} = {\cal T}^{(1)} + {\cal T}^{(2)} 
+{\cal T}^{(3)}$, with components defined 
(using operator notation for compactness) as
\begin{equation}\label{TFINAL}
{\cal T}^{(k)} = {\cal T}_{ij}+ {\cal T}_{ij}
\frac{Q_3}
{\Omega-\epsilon_1-\epsilon_2-\epsilon_3+i\eta}
 \left({\cal T}^{(i)}+ {\cal T}^{(j)}\right).
\end{equation}
These are nonsingular type II Fredholm integral equations; the operator 
$Q_3$ in momentum representation is given by 
$Q_{3}(\mathbf k_1,\mathbf k_2 ,\mathbf k_3 ) =
[1-f(\mathbf k_1)][1-f(\mathbf k_2)][1-f(\mathbf k_3)]-
f(\mathbf k_1)f(\mathbf k_2)f(\mathbf k_3)$; and $\epsilon_i(\mathbf k_i)$
are the quasiparticle spectra. 
The momentum space for the three-body problem is conveniently 
spanned by the Jacobi four-momenta $K = k_i+ k_j+ k_k$, $p_{ij} 
= (k_i-k_j)/2$, and $q_{k} = (k_i+k_j)/3 -2k_k/3$. 
The ${\cal T}_{ij}$-matrices are essentially the two-body 
scattering amplitudes, embedded in the Hilbert space of 
three-body states.  In the momentum representation they are determined 
from
\begin{equation}
\langle p \vert T(\omega)\vert p'\rangle = 
\langle p \vert V\vert p'\rangle + \int\!\!\frac{ dp'' {p'}'^2}{4\pi2}
\langle p' \vert V\vert p''\rangle
\frac{ Q_2(p,q)}{\omega-\epsilon_+(q,p)-\epsilon_-(q,p)+i\eta }
\langle p'' \vert T(\omega)\vert p'\rangle ,
\end{equation}
where $Q_2(q,p)= \langle 1-f(\mathbf q/2+\mathbf p)-f(\mathbf q/2-\mathbf p)\rangle$ 
and $\epsilon_{\pm }(q,p) = \langle \epsilon(\mathbf q/2\pm \mathbf p)\rangle$
are averaged over the angle between the vectors $\mathbf q$ and $\mathbf p$.   
\begin{figure}[tb]
\begin{center}
\epsfig{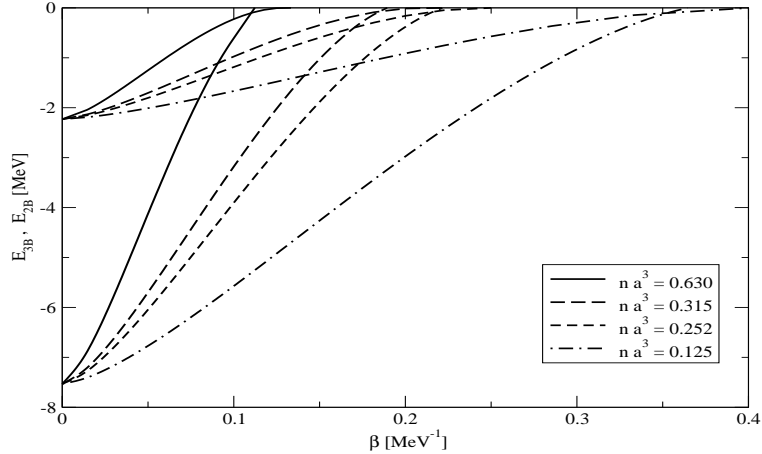}
\caption{
Dependence of the two-body ($E_d$) and three-body ($E_t$)
binding energies on inverse temperature, for fixed values of 
the ratio $f = n_0/n$, where $n$ is the baryon density and
$n_0 = 0.16$ fm$^{-3}$ is saturation density of nuclear matter. 
For asymptotically large temperature, $E_d(\infty) = -2.23$ MeV
and $E_t(\infty) = -7.53$ MeV. The ratio $E_t(\beta)/E_d(\beta)$
is a universal constant independent of temperature~\cite{Sedrakian:2005db}.
 }
\label{fig:2B3B}
\end{center}
\end{figure}
Compared to the free-space problem, the three-body equations in the 
background medium now include two- and three-body propagators 
that account for (i)~the suppression of the 
phase-space available for scattering in intermediate
two-body states, encoded in the functions
$Q_{2}$, (ii)~the phase-space occupation for
the intermediate three-body states, encoded in the function
$Q_{3}$, and (iii)~renormalization of the single-particle energies 
$\epsilon(p)$.  For small temperatures the quantum degeneracy is large
and the first two factors significantly reduce  
the binding energy of a three-body bound state; at a 
critical temperature $T_{c3}$ corresponding to $E_t(\beta) = 0$,
the bound state enters the continuum. 

This behavior is illustrated in Fig.~\ref{fig:2B3B}, which shows
the temperature dependence of the two- and three-body bound-state energies
in dilute nuclear matter for several values of the density of
the environment.   In analogy to the behavior of the
in-medium three-body bound state, the binding energy of the two-body 
bound state enters the continuum at a critical temperature $T_{c2}$,
corresponding to the condition $E_d(\beta) = 0$.  
Our solutions exhibit a remarkable feature:
the ratio $\eta = E_t(\beta)/E_d(\beta)$
is a constant {\it independent of temperature}.
For the chosen potentials, the asymptotic free-space values
of the binding energies are $E_t(0) = -7.53$ MeV and $E_d(0) = -2.23$ MeV;
hence $\eta = 3.38$.  An alternative definition of the critical
temperature for trimer extinction is $E_t(\beta_{3c}') = E_d(\beta)$.
This definition takes into account the break-up channel
$t\to d+ n$ of the three-body bound state into the two-body
bound state $d$ and a nucleon $n$.  The difference between the two
definitions is numerically insignificant.

\begin{figure}[hbt]%

\begin{center}
\hspace{-2cm}
 \parbox{2.2in}{\epsfig{figure=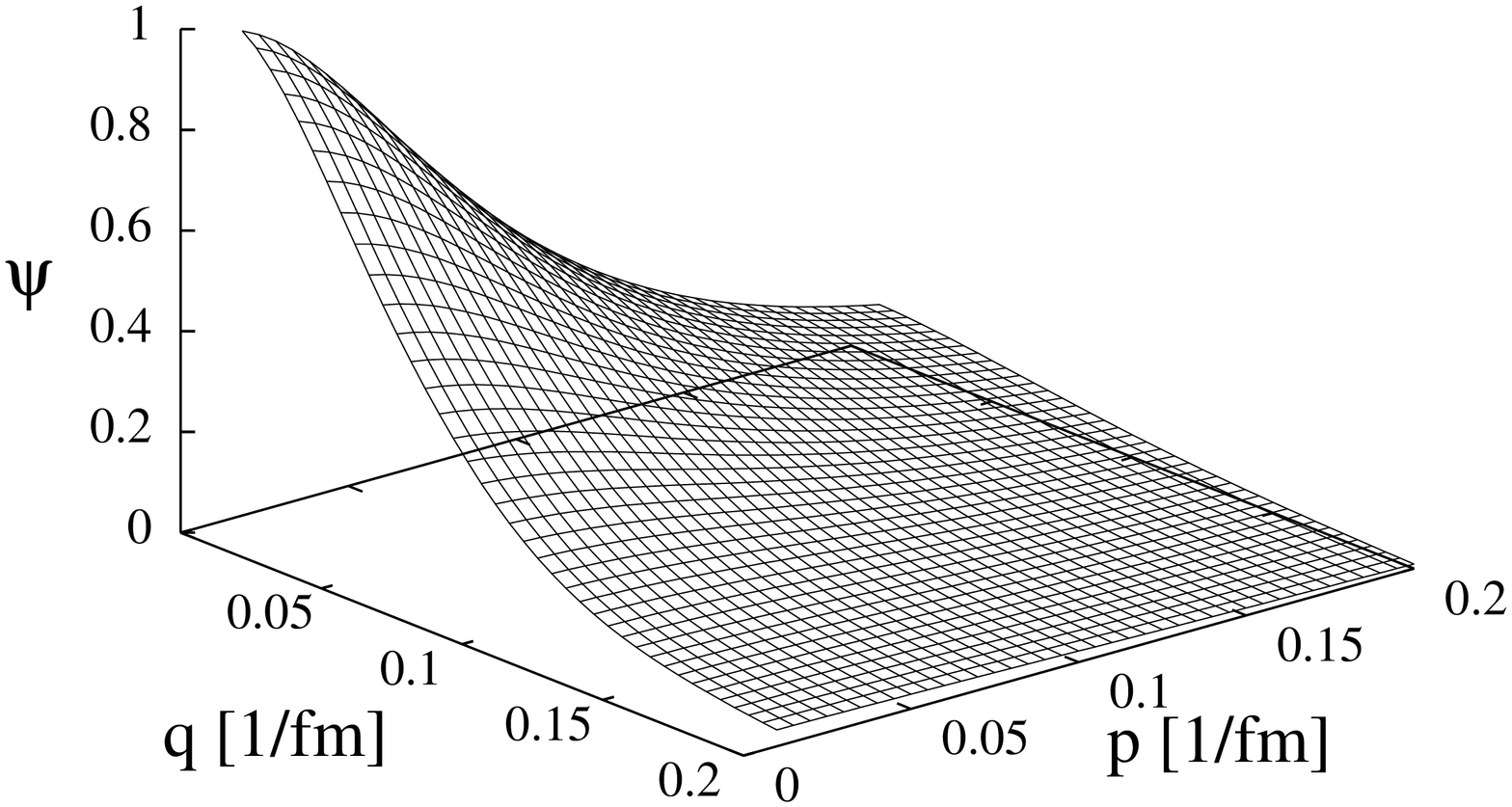,width=2.0in,height=2.8in,angle=-90}
 }
 \hspace*{.8cm}
 \parbox{2.2in}{\epsfig{figure=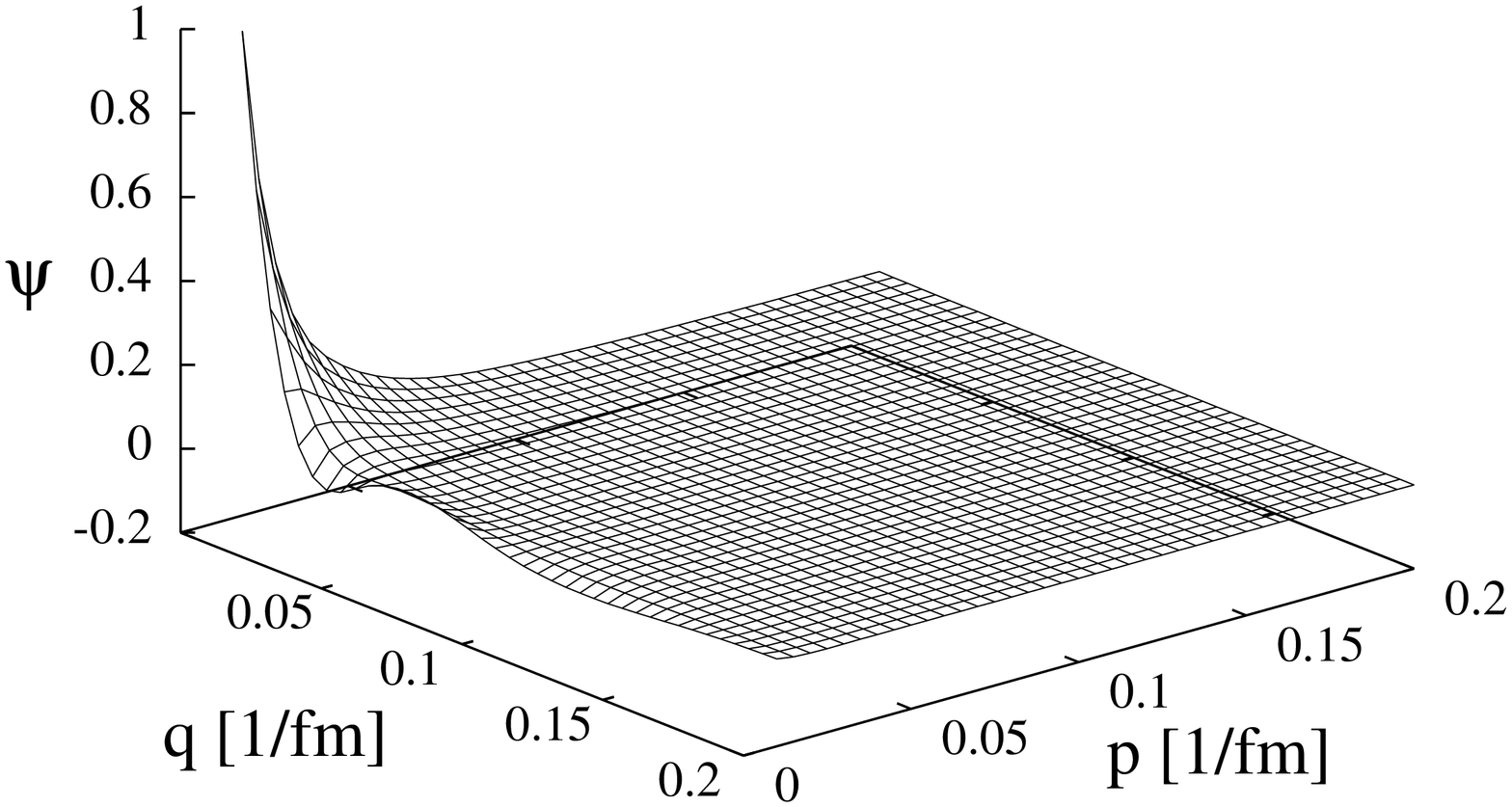,width=2.0in,height=2.8in,angle=-90}
 }
 \caption{Wave function of the three-body bound state as a function 
of the Jacobi momenta $p$ and $q$ defined in the text, for
$f = n_0/n = 60$ and temperatures $T = 60$ (left panel) 
and 6.6 MeV (right panel).
}
\label{fig:WF}
\end{center}
\end{figure}

Fig.~\ref{fig:WF} depicts the normalized three-body bound-state wave
function for three representative temperatures, as a function of the 
Jacobi momenta $p$ and $q$. As the temperature drops, the wave 
function becomes increasingly 
localized around the origin in momentum space.  Correspondingly, the 
radius of the bound state increases in $r$-space, eventually tending 
to infinity at the transition.  The wave-function oscillates near
the transition temperature (right panel of Fig.~\ref{fig:WF}). 
This oscillatory behavior is a precursor of the transition to the 
continuum, which in the absence of a trimer-trimer interaction 
is characterized by plane-wave states.

\section{The weak interaction in cold subnuclear matter}

%
Non-nucleonic channels of cooling that operate in the crusts of neutron 
stars are electron neutrino bremsstrahlung off nuclei 
and plasmon decay~\cite{plasmon}:
$
e+(A,Z) \to e+(A,Z)+\nu+\bar\nu$ and $ {\rm plasmon}\to +\nu+\bar\nu.
$
%
Above the critical temperature $T_c$ for neutron superfluidity, 
the neutrons that occupy
continuum states (i.e., those not bound in clusters) 
emit neutrinos of all flavors $f$ via the bremsstrahlung 
process\cite{Maxwell} $n+n \to n+n+\nu_f+\bar\nu_f$.
At $T \le T_c$ the latter process is suppressed exponentially by
${\rm exp}(-2\Delta/T)$, where $\Delta$ is the gap in the quasiparticle 
spectrum. The superfluid nature of the matter allows for a 
neutrino-generating reaction
(known as pair-breaking and recombination), whose rate scales like
$\Delta^7$ and thus is specific to the superfluid (i.e., vanishes as
$\Delta\to 0$). The rate of the process is given by the 
polarization tensor of superfluid matter~\cite{PB}. 

A systematic diagrammatic method to compute the reaction rates is
based on the
kinetic equation for neutrino transport, formulated in terms of 
real-time Green's functions~\cite{KIN}. 
The corresponding Boltzmann equation is 
\begin{equation}
\left[\partial_t + {\mathbf \partial}_q\,\omega_{\nu} 
(\mathbf q) \mathbf\partial_x
\right] f_{\nu}({\mathbf q},x)
\int_{0}^\infty \frac{dq_0}{2\pi} 
{\rm Tr} \left[\Omega^<(q,x)S_0^>(q,x)-
\Omega^>(q,x)S_0^<(q,x)\right],
\nonumber 
\end{equation}
where $q \equiv (q_0,{\bf q})$ is the four momentum, 
$S_0^{>,<}(q,x)$ are the neutrino propagators, and
$\Omega^{>,<}(q,x)$ are their self-energies.   In second Born 
approximation with respect to the weak vertices $\Gamma_{L\, q_1}^{\mu}$,
the latter
are given in terms of the polarization tensor(s) 
$ \Pi^{>,<}_{\mu\lambda}(q_1,x)$ 
of ambient matter as
\begin{equation}
-i\Omega^{>,<}(q,x)= \int \frac{d4 q_1}{(2\pi)^4}
\frac{d4 q_2}{(2\pi)^4}(2\pi)^4 \delta^4(q - q_2 - q_1)
i\Gamma_{L\, q_1}^{\mu}\, iS_0^{<}(q_2,x) i\Gamma_{L\, q_1}^{\dagger\, \lambda}
i \Pi^{>,<}_{\mu\lambda}(q_1,x),
\end{equation}
The ``greater'' and ``lesser'' signs refer to the ordering of 
two-point functions along the Schwinger contour in the standard way.

It is the total loss of energy in neutrinos per unit time and 
unit volume, i.e., the emissivity, that is of interest for
the astrophysics of compact stars.  This quantity is obtained 
by integrating the first moment of Boltzmann equation. For the 
bremsstrahlung of neutrinos and anti-neutrinos of given flavor 
it is expressed as
\begin{eqnarray}\label{EMISSIVITY}
\epsilon_{\nu\bar\nu} = -\frac{G^2}{4}
\sum_{q_1,q_2}
\int\! d^4 q \,\delta (q_1 + q_2 -  q)q_0
 g(q_0) \Lambda^{\mu\zeta}(q_1,q_2){\rm Im}\,\Pi_{\mu\zeta}(q),
\end{eqnarray}
where $G$ is the weak coupling constant, $q$ is the four-momentum transfer,
$g(q_0) = [{\rm exp}(q_0/T) - 1]^{-1}$ is the Bose distribution 
function, $\Pi_{\mu\zeta}(q)$ is the retarded polarization tensor, 
and $\Lambda^{\mu\lambda}(q_1,q_2) = {\rm Tr}\left[\gamma^{\mu}
(1 - \gamma5)\!\not\! q_1\gamma^{\nu}(1-\gamma5)\!\not\! q_2\right]$. 
Sums over the neutrino momenta $q_{1,2}$ indicate integration over 
the invariant phase-space volume. The central problem of the theory
is to compute the polarization tensor of the cold subnuclear matter. 
Initially,  calculations of the polarization tensor within the superfluid 
phase were carried out 
at the one-loop approximation. This treatment was recently shown to be 
inadequate for the $S$-wave superfluid in neutron-star 
crusts~\cite{Sedrakian:2006ys}.
\begin{figure}[t]
\begin{center}
\includegraphics[height=1.5cm,width=12cm]{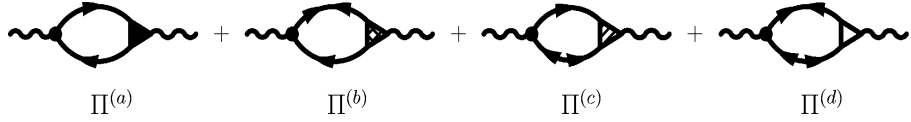}
\end{center}
\caption[]               
{The sum of polarization tensors that
 contribute to the neutrino emission rate.  The contributions
form $\Pi^{(b)}(q)$ and $\Pi^{(c)}(q)$ vanish at the one-loop
approximation.
}
\label{fig:pol_functions}
\end{figure}

A many-body framework that is consistent with the sum rules for 
the polarization tensor, in  particular with the $f$ sum rule 
\begin{eqnarray}
\lim_{\mathbf q\to 0}\int  
d\omega ~\omega ~{\rm Im}\Pi^V(\mathbf q,\omega) = 0,
\end{eqnarray}
is provided by the random-phase resummation of the particle-hole 
diagrams in the superfluid matter. Because of the Nambu-Gorkov extension 
of the number of possible propagators in the superfluid phase, which now
include both normal ($G$) and anomalous ($F$) ones,
at least three topologically different vertices are involved,
which obey (schematically) the following equations 
\begin{eqnarray}\label{g1}
\hat\Gamma_1  &=& \Gamma_0  
             +  v  (G\Gamma_1 G
             +       \hat F\Gamma_3 G
             +       G\Gamma_2 \hat F
             +       \hat F\Gamma_4 \hat F),  \\ 
\label{g2}
\hat\Gamma_2  &=&  \quad \quad v  (G\Gamma_2 G^{\dagger}
             +       \hat F\Gamma_4 G^{\dagger}
             +       G\Gamma_1 \hat F
             +       \hat F\Gamma_3 \hat F), \\ 
\label{g3} 
\hat\Gamma_3  &=&  \quad \quad v  (G^{\dagger}\Gamma_3 G
             +       \hat F\Gamma_1 G
             +       G^{\dagger}\Gamma_4 \hat F
             +       \hat F\Gamma_2 \hat F),
\end{eqnarray}
$v$ being the scalar interaction in the particle-hole channel.
The fourth integral equation for the vertex $\Gamma_4 $ follows 
upon interchanging particle and hole propagators in Eq.~(\ref{g1}).
The full polarization tensor is the sum 
of the contributions shown in Fig.~\ref{fig:pol_functions}. 
It can be expressed through ``rotated'' polarization functions
${\cal A, B,}$ and ${\cal C}$ as
\begin{eqnarray} \label{FULL_PI}
\Pi^V (q) = \frac{{\mathcal A}(q){\mathcal C}(q)+{\mathcal B}(q)2}
{{\mathcal C}(q) -v^V[{\mathcal A}(q){\mathcal C}(q)+{\mathcal B}(q)2]}
\end{eqnarray}
with
${\cal A} = 2\Delta^2~I_0(q)-\Delta^2\xi_q ~I_A(q)$,
${\cal B} = - \omega\Delta~I_0(q)$, and
${\cal C}(q) = - ({\omega^2}/{2})~I_0(q)
+\xi_q~I_C(q)$, where $\xi_q= q^2/2m$ is the nucleon recoil.
(The integrals $I_0$, $I_C$, and $I_A$ can be found in 
Ref.~\citen{Sedrakian:2006ys}.)
It is now manifest that  $\Pi^V (q) = 0 $ when $q=0$. Thus, the leading
order contribution to the polarization tensor appears at $O(q^2)$ 
and is linear in $\xi_q$. Since the neutrinos
are thermal, with energies $\omega \sim \vert {\bf q} \vert \sim T$, 
the polarization tensor is suppressed by a factor $T/m$, which is of 
order $5\times 10^{-3}$.

The emissivity of the pair-breaking process can be compared to that of 
the modified bremsstrahlung (MB) process $n+n \to n+n+\nu+\bar\nu$, 
which is suppressed by roughly a factor ${\rm exp}(-2\Delta/T)$
in the superfluid phase. Thus, the ratio of the neutrino 
loss rate through MB to that from the pair-breaking process, 
as computed by Friman and Maxwell,\cite{Maxwell} is
\begin{equation}
R = \frac{2460 \pi^4}{14175}\kappa\left(\frac{g_A}{c_V}\right)^2
\left(\frac{T}{\Delta}\right)^2\left(\frac{m_n^*}{m_{\pi}}\right)^4 
\frac{F}{I_0}{\rm exp}\left(-\frac{2\Delta}{T}\right),
\quad I_0 = \int_{\frac{2\Delta}{T}}^{\infty} dx x^5 
f\left(\frac{x}{2}\right)^2,\nonumber
\end{equation}
where $g_A$ and $c_V$ are the weak axial and vector coupling constants, 
$m_n^*$ and $m_{\pi}$ are the neutron and pion masses, and
$F\simeq 0.6$ (defined in Ref.~\citen{Maxwell}).  The factor
$\kappa = 0.2$ accounts for the correction to the one-pion-exchange 
rate due to the full resummation of ladder series in neutron 
matter. The pair-breaking process dominates the MB process 
for temperatures below $0.8T_c$, where it is most efficient.
This is illustrated in the table above.
\begin{table}[t]
\begin{tabular}{ccccccc}
\hline
$k_F$ [fm$^{-1}$] & $m^*/m$ & $\Delta$ [MeV] & $T_c$ [MeV] & $R(0.5)$ &
$R(0.8)$ &  $R(0.9)$\\
\hline
0.8 & 0.97 & 3.15 & 1.78 & 0.014 & 1.0 & 6.5\\
1.6 & 0.84 & 0.57 & 0.38 & 0.022 & 1.0 & 7.4\\
\hline
\end{tabular}
\begin{tabnote}
Quoted are the wave number of 
neutrons, their effective mass, the gap and critical temperature, and 
the ratio $R$ as a function of reduced temperature $T/T_c$.
\end{tabnote}
\end{table}
The comparison made here should be taken with caution, since the exponential 
suppression of the MB rate is not accurate (within a factor of a few) 
at temperatures close to the critical temperature. Nevertheless, one may 
safely conclude that the vector-current pair-breaking process is competitive 
with the modified pair-bremsstrahlung process in the relevant temperature 
domain $T/T_c\in [0.2-1]$.

\section{Closing remarks}
Subnuclear matter at finite temperatures offers a fascinating arena
for the development of many-body theory. Since the interactions 
invoved are well constrained by experiment, the entire complexity 
arises from the many-body correlations. As shown in the examples 
chosen, subnuclear 
matter may exhibit a range of salient many-body phenomena, such as 
Bose-Einstein condensation of alpha particles, BEC-BCS crossover in the 
deuteron channel, many-body extinction of bound states with increasing 
degeneracy, and non-trivial and quantitatively important modifications of 
the weak interaction rates due to many-body effects.

\section*{Acknowledgements}
We thank H. M\"uther and P. Schuck for their contribution to the research 
described in this article. We are grateful to the organizers of RPMBT14 
for their impressive efforts and dedication in arranging a most
successful conference.


\begin{thebibliography}{99}



\bibitem{Baym:1999ws} 
G.  Baym, J.-P. Blaizot, and J. Zinn-Justin, {\it Europhys. Lett.}
      {\bf 49} (2),  150 (2000).

\bibitem{ZinnJustin:2000dr}
  J.~Zinn-Justin,
  arXiv:hep-ph/0005272.


\bibitem{Sedrakian:2004fh}
  A.~Sedrakian, H.~M\"uther and P.~Schuck,
{\it   Nucl. Phys.} A 766,  97-106 (2006).


\bibitem{ALPHA} 
M. T. Johnson and J. W. Clark, {\it Kinam} 
{\bf 3}, 3 (1980) also made available 
at this URL {\tt http://wuphys.wustl.edu/Fac/facDisplay.php?name=Clark.txt}


\bibitem{NSR} P. Nozi\`eres and S. Schmitt-Rink, 
     {\it J. Low Temp. Phys.} {\bf 59},   195 (1985).


\bibitem{Alm:1991ne}
  T.~Alm, B.~L.~Friman, G.~R\"opke and H.~Schulz,
  {\it  Nucl. Phys.} {\bf A 551}, 45 (1993).

\bibitem{Lombardo:2001ek}
  U.~Lombardo, P.~Nozi\`eres, P.~Schuck, H.~J.~Schulze and A.~Sedrakian,
 {\it  Phys.\ Rev.\ } C {\bf 64}, 064314 (2001)
  [arXiv:nucl-th/0109024].



\bibitem{Sedrakian:2005db}
  A.~Sedrakian and J.~W.~Clark,
{\it  Phys.\ Rev.\ }  C {\bf 73},  035803 (2006).

\bibitem{Sedrakian:2006xm}  A. Sedrakian and J. W. Clark, in
{\it "Pairing in Fermionic Systems: 
Basic Concepts and Modern Applications",} eds. A. Sedrakian, J. W. Clark, 
and M. Alford, World Scientific, pp. 145-175, 
 [arXiv:nucl-th/0607028].



\bibitem{plasmon} G.~G. Festa and M. A. Ruderman, 
{\it Phys. Rev.} {\bf 122},
                      1317 (1969); 
J. B. Adams, M. A. Ruderman, and C. H. Woo, 
                      {\it Phys. Rev.} {\bf 129}, 1383 (1963).

\bibitem{Maxwell} O. V. Maxwell and B. L. Friman, 
{\it Astrophys. J.} {\bf 232}, 541 
(1979);  D.~N.~Voskresensky and A.~V.~Senatorov, 
{\it Sov. J. Nucl. Phys.} 
{\bf 45},  411 (1987)
[{\it Yad. Fiz.} {\bf 45}, 657 (1987)]; A. Sedrakian 
and A. E. L. Dieperink, {\it Phys. Lett.} B {\bf 463};
 E.  van Dalen, A. E. L. Dieperink, and J. A. Tjon,
{\it Phys. Rev.} C {\bf 67}, 580 (2003).


\bibitem{PB}  E.~G.~Flowers, M.~Ruderman, and P.~G.~Sutherland,
                   {\it  Astrophys. J.} {\bf 205}, 541 (1976); 
                D.~N.~Voskresensky and A.~V.~Senatorov,
                    {\it Sov. J. Nucl. Phys.} {\bf 45},  411 (1987)
                    [{\it Yad. Fiz.} {\bf 45}, 657 (1987)].
                A. B. Migdal, E. E. Saperstein, M. A. Troitsky, 
                and D. N.  Voskresensky, {\it Phys. Rep.} 
{\bf 192},  179 (1990);

\bibitem{KIN} A. Sedrakian 
and A. E. L. Dieperink,  {\it Phys. Rev. }
D  {\bf 62},  083002 (2000); A.~Sedrakian, arXiv:astro-ph/0701017;
see also  D.~N.~Voskresensky and A. V. Senatorov in ref. ~\refcite{PB}.



\bibitem{Sedrakian:2006ys}
  A.~Sedrakian, H.~M\"uther and P.~Schuck,
  arXiv:astro-ph/0611676; 
see also 
  L.~B.~Leinson and A.~Perez,
  {\it Phys.\ Lett.\ } B {\bf 638}, 114 (2006)
  [arXiv:astro-ph/0606651].

\end{thebibliography}
\end{document}